\documentclass[journal=nalefd,manuscript=Letter,layout=twocolumn]{achemso}

\usepackage{chemformula} 
\usepackage[T1]{fontenc} 
\usepackage[latin9,utf8]{inputenc}
\usepackage{float}
\usepackage{textcomp}
\usepackage{amsmath}
\usepackage{graphicx}
\usepackage{tabularx}
\usepackage{booktabs}
\usepackage{multirow}
\usepackage{url}
\usepackage{grffile}
\usepackage{bbold}
\usepackage{dcolumn}
\usepackage{color}
\usepackage{xcolor}
\DeclareGraphicsExtensions{.png .jpg .pdf}
\usepackage{hyperref}
\usepackage{braket}
\usepackage{physics}
\usepackage{array}
\usepackage{booktabs}
\usepackage{soul}
\usepackage{lipsum}

\hypersetup{
    colorlinks,
    linkcolor={blue!50!black},
    citecolor={blue!50!black},
    urlcolor={blue!80!black}
}

\makeatother

\author{C. Serati de Brito}
\affiliation{Physics Department, Federal University of São Carlos, São Carlos, SP, 13565-905, Brazil}
\alsoaffiliation{Institut für Experimentelle und Angewandte Physik, Universit\"at Regensburg, D-93040 Regensburg, Germany}

\author{P. E. {Faria~Junior}}
\affiliation{Institute of Theoretical Physics, University of Regensburg, 93040 Regensburg, Germany}

\author{T. S. Ghiasi}
\affiliation{Kavli Institute of Nanoscience, Delft University of Technology, Lorentzweg 1, 2628 CJ Delft, The Netherlands}

\author{J. Ingla-Aynés}
\affiliation{Kavli Institute of Nanoscience, Delft University of Technology, Lorentzweg 1, 2628 CJ Delft, The Netherlands}

\author{C. R. Rabahi}
\affiliation{Physics Department, Federal University of São Carlos, São Carlos, SP, 13565-905, Brazil}

\author{C. Cavalini}
\affiliation{Physics Department, Federal University of São Carlos, São Carlos, SP, 13565-905, Brazil}

\author{F. Dirnberger}
\affiliation{Institute of Applied Physics and Würzburg-Dresden Cluster of Excellence ct.qmat,
Technische Universität Dresden, Germany}

\author{S. Mañas-Valero}
\affiliation{Kavli Institute of Nanoscience, Delft University of Technology, Lorentzweg 1, 2628 CJ Delft, The Netherlands}
\alsoaffiliation{Instituto de Ciencia Molecular (ICMol), Universitat de València, Catedrático José Beltrán 2, Paterna 46980, Spain}

\author{K. Watanabe}
\affiliation{Research Center for Materials Nanoarchitectonics, National Institute for Materials Science,  1-1 Namiki, Tsukuba 305-0044, Japan}

\author{T. Taniguchi}
\affiliation{Research Center for Materials Nanoarchitectonics, National Institute for Materials Science,  1-1 Namiki, Tsukuba 305-0044, Japan}

\author{K. Zollner}
\affiliation{Institute of Theoretical Physics, University of Regensburg, 93040 Regensburg, Germany}

\author{J. Fabian}
\affiliation{Institute of Theoretical Physics, University of Regensburg, 93040 Regensburg, Germany}

\author{C. Sch\"uller}
\affiliation{Institut für Experimentelle und Angewandte Physik, Universit\"at Regensburg, D-93040 Regensburg, Germany}

\author{H. S. J. van der Zant}
\affiliation{Kavli Institute of Nanoscience, Delft University of Technology, Lorentzweg 1, 2628 CJ Delft, The Netherlands}

\author{Y. Galvão Gobato}
\email{yara@df.ufscar.br}
\affiliation{Physics Department, Federal University of São Carlos, São Carlos, SP, 13565-905, Brazil}

\title{Charge transfer and asymmetric coupling of MoSe$_2$ valleys to the magnetic order of CrSBr}

\keywords{Transition Metal Dichalcogenides, two-dimensional magnets, van der Waals Heterostructures, Proximity Effects, Magneto-Optics.}

\begin{document}





\begin{abstract}
Van der Waals (vdW) heterostructures composed of two-dimensional (2D) transition metal dichalcogenides (TMD) and vdW magnetic materials offer an intriguing platform to functionalize valley and excitonic properties in non-magnetic TMDs. Here, we report magneto-photoluminescence (PL) investigations of monolayer (ML) MoSe$_2$ on the layered A-type antiferromagnetic (AFM) semiconductor CrSBr under different magnetic field orientations. Our results reveal a clear influence of the CrSBr magnetic order on the optical properties of MoSe$_2$, such as an anomalous linear-polarization dependence, changes of the exciton/trion energies, a magnetic-field dependence of the PL intensities, and a valley $g$-factor with signatures of an asymmetric magnetic proximity interaction. Furthermore, first principles calculations suggest that MoSe$_2$/CrSBr forms a broken-gap (type-III) band alignment, facilitating charge transfer processes. 
The work establishes that antiferromagnetic-nonmagnetic interfaces can be used to control the valley and excitonic properties of TMDs, relevant for the development of opto-spintronics devices.
\end{abstract}

\section{}

Recently, van der Waals (vdW) magnetic materials have attracted  increasing attention because of their unique magnetic properties and possible applications in spintronics\cite{Lyons2020, Zhong2020, Onga2020, Li2022, Choi2022, Seyler2018, Pawbake2023, Scharf2017, Huang2017, Zhang2019, Xie2018, Ahn2020, Boix-Constant2022}. Several studies were performed in heterostructures using magnetic materials and  monolayer TMDs\cite{Geim2013,Novoselov2016,Choi2022_MoSe2-CrBr3, Onga2020, Zhong2017, Zhong2020, Lyons2020, Huang2020,Mak2019, Zollner2019, Zollner2020PRB, Zollner2023PRB}. These heterostructures employ magnetic proximity effects to modify the physical properties of the ML TMD adjacent to the magnetic material and therefore offer new opportunities for engineering magnetic heterostructures\cite{Huang2020}. Actually, recent studies evidenced an enhanced valley splitting of WSe$_2$ and WS$_2$ monolayers on the ferromagnetic (FM) material EuS\cite{Zhao2017,Norden2019}, a giant zero-field valley splitting of MoSe$_2$/CrBr$_3$\cite{Ciorciaro2020}, asymmetric magnetic proximity interactions in MoSe$_2$/CrBr$_3$\cite{Choi2022_MoSe2-CrBr3}, and an anomalous temperature dependence of the MoSe$_2$/MnPSe$_3$ excitonic peak below the Néel temperature ($T_N$) \cite{Onga2020}. Furthermore, magnetic proximity effects have led to spin-dependent charge transfer and concomitant circularly polarized PL in hybrid devices based on both CrI$_3$\cite{Zhong2017,Zhong2020} and CrBr$_3$\cite{Lyons2020}. However, most previous studies in magnetic vdW heterointerfaces involved vdW ferromagnetic materials\cite{Lyons2020,Zhong2020,Onga2020,Li2022,Choi2022,Seyler2018}. AFM materials have a variety of spin orderings with distinct magnetic symmetry groups which could result in unique magnetic properties and therefore  there are interesting ways to control their functionalities by choosing appropriate AFM materials\cite{Onga2020}.

In this work, we investigate the impact of the CrSBr antiferromagnetic substrate on the exciton and valley properties of ML MoSe$_2$. We have performed micro-PL measurements under a magnetic field along each of the three crystallographic axes of CrSBr.  In general, our findings show that the exciton and valley properties of ML TMDs can be engineered by the interplay of magnetic proximity, efficient charge transfer effects, exciton/trion-magnon coupling and dielectric anomalies of 2D antiferromagnetic materials.

The layered magnetic material CrSBr is a vdW direct gap semiconductor with A-type AFM and Neél temperature of~132~K in its bulk form.\cite{Klein2023,Lee2021,Wilson2021,Telford2022,Lopez-Paz2022,Ye2022,Ghiasi2021,Bae2022,Klein2023-2}. In addition, CrSBr presents another phase transition around the temperature of $T = 40$~K \cite{Pawbake2023,Telford2022,Lopez-Paz2022} which is not well understood but might be related to crystal defects \cite{Klein2023} or  spin-freezing effects\cite{Boix-Constant2022}. The CrSBr crystal consists of layers with rectangular unit cells in the plane ($\hat{a}$ - $\hat{b}$) which are stacked along the $\hat{c}$ axis to produce an orthorhombic structure [Figure 1(b)]. The optical properties of CrSBr reflect its highly anisotropic electronic and magnetic structure. A prominent example is the coupling of excitons to the magnetic order. Changes in the static magnetic configuration induced by applying a magnetic field, for instance, directly impact the exciton energy. The electronic band structure, and consequently the energy of excitons in CrSBr, are sensitive to the interlayer magnetic exchange interaction which can be used to probe its magnetic properties\cite{Pawbake2023,Wilson2021,Lopez-Paz2022,Klein2023-2}. 

Monolayer MoSe$_2$ is a direct band gap semiconductor with two inequivalent $\pm$K valleys and robust excitons \cite{Mak2010,Splendiani2010,Xu2014,Xiao2012,Mak2012,Zeng2012,Schaibley2016,Gobato2022,Ishii_2019}. Under out-of-plane magnetic fields, valley Zeeman effects and magnetic-field-induced valley polarization are observed and these effects  depend on the presence of strain, doping and magnetic proximity effects\cite{Aivazian2015, Li2014, Srivastava2015, Macneill2015, Wang2015, Mitioglu2015, Stier2016, Plechinger2016, Covre2022,Wozniak2020,Deilmann2020PRL,Xuan2021npj,Back2017,Ciorciaro2020}.

\begin{figure*}[ht]
\centering{\includegraphics[width=2.0\columnwidth]{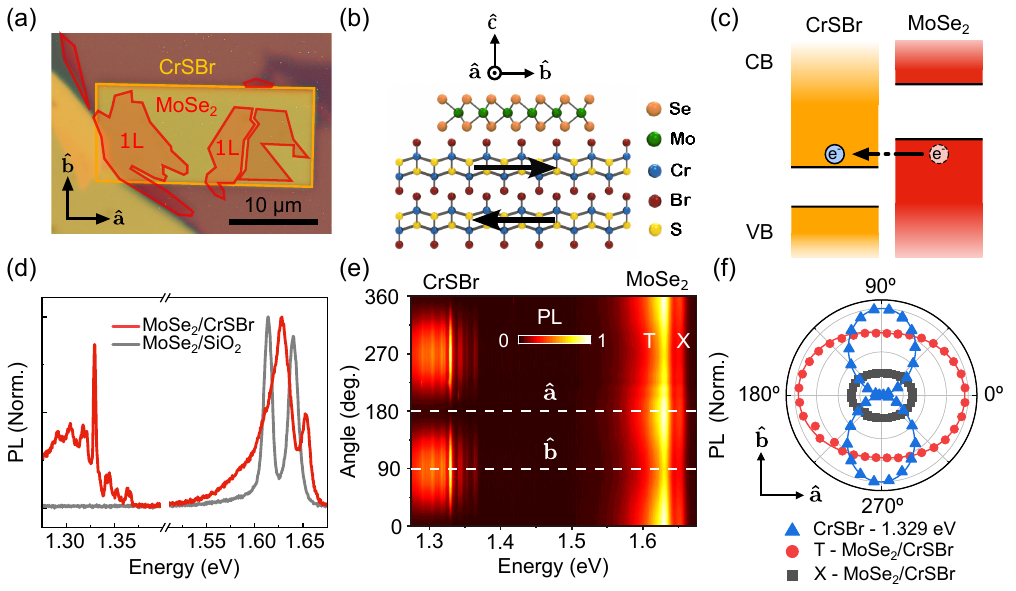}}
\caption{(a) Optical microscope image of the studied ML MoSe$_2$/bulk CrSBr vdW heterostructure, covered with a thin layer of hBN. The thickness of our CrSBr layer is about 35~nm (Figure~S1) (b) Schematics of the crystal structure of the heterostructure. (c) Schematics of the band alignment of ML MoSe$_2$/bulk CrSBr and charge transfer from the MoSe$_2$ valence band (VB) to the CrSBr conduction band (CB). (d) Typical PL spectra from ML MoSe$_2$/CrSBr and  MoSe$_2$/SiO$_2$ regions at 3.6~K.  The laser energy is 1.88~eV. (e) Color-coded map of the linearly-polarized emission intensity as a function of the angle of in-plane polarization. The laser excitation is linearly polarized along the $\hat{a}$ axis (f) Polar plot of the PL intensity versus the in-plane linear polarization angle for the most intense PL peak energy of CrSBr (1.329~eV) and  also for the exciton (X)  and trion (T) emission peaks from MoSe$_2$ on CrSBr.}
\label{fig.LinearDependence}
\end{figure*}

 Figure~\ref{fig.LinearDependence}~(a) shows an optical microscope image of our MoSe$_2$/CrSBr heterostructure and the crystal orientations, $\hat{a}$ and $\hat{b}$, of the CrSBr bulk crystal, while in Figure~\ref{fig.LinearDependence}~(b) the MoSe$_2$ and CrSBr crystal structures are sketched. Figure~\ref{fig.LinearDependence}~(c) presents the predicted type-III (broken-gap) band alignment of the heterostructure. In Figure~\ref{fig.LinearDependence}~(d), the PL spectrum of CrSBr at 3.6~K is displayed in the left part. Several PL peaks are observed below 1.4 eV  and associated with excitons\cite{Pawbake2023,Klein2023-2, Dirnberger2023,Marques-Moros2023}, defects\cite{Klein2023}, and strong exciton-photon coupling \cite{Lin2023}.

Figure~\ref{fig.LinearDependence}~(d) shows the exciton and trion peaks in the normalized PL spectra of MoSe$_2$/CrSBr and  MoSe$_2$/SiO$_2$. According to our theoretical predictions for the band alignment [see Figures 2~(e-f)], the MoSe$_2$ layer may be strongly p-doped, while MoSe$_2$ on SiO$_2$ is usually n-doped\cite{Covre2022,Macneill2015}. Therefore, the trion in MoSe$_2$/CrSBr is most likely a positively charged exciton.  In addition, a low-energy shoulder in the trion emission is observed which could be due to inelastic scattering of trions by magnons, or due to localized trions at charged impurities \cite{Schwemmer2017}.

\begin{figure*}[htb]
\centering{\includegraphics[width=2.0\columnwidth]{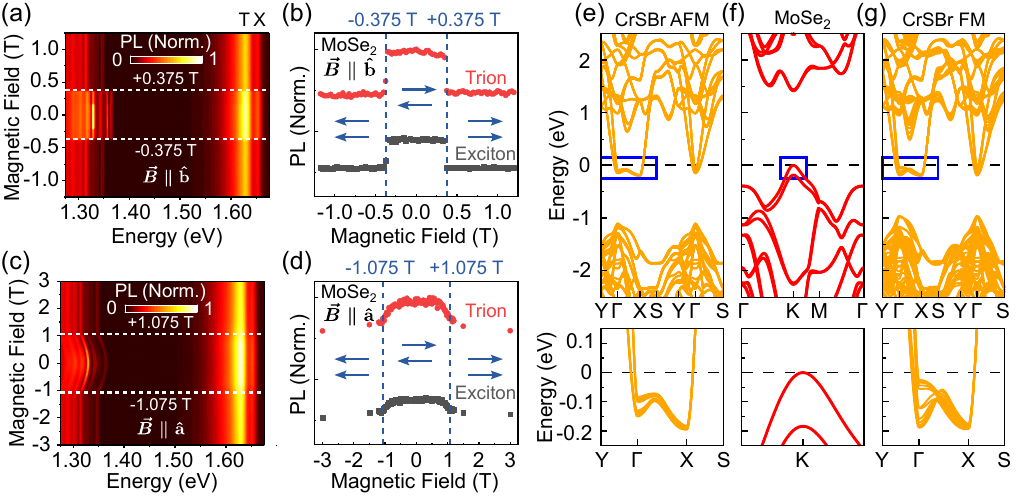}}
\caption{(a,c) Color-code map for circularly-polarized PL intensity from the MoSe$_2$/CrSBr heterostructure as a function of the in-plane magnetic field, oriented along the in-plane easy (for $\Vec{B} \parallel \hat{b}$) and intermediate axes (for $\Vec{B} \parallel \hat{a}$). The excitation is performed using a linearly polarized laser. The PL detection is $\sigma^{-}$ for positive magnetic fields. (b,d) Magnetic-field  dependence of the MoSe$_2$  PL intensity of the exciton and trion emissions for both field orientations; the MoSe$_2$ PL intensity is sensitive to the magnetic phases of the CrSBr. Calculated band structure with spin-orbit coupling for (e) CrSBr AFM, (f) ML MoSe$_2$, and (g) CrSBr FM systems. The CrSBr systems consist of 6 layers. The horizontal dashed lines indicate the Fermi energy aligned with respect to the vacuum levels. The bottom panels show the band structure in the region of the blue rectangles, indicating clear differences of the AFM and FM energy levels next to the Fermi energy.}
\label{fig.BField_Parallel+BS}
\end{figure*}

Next, we investigate the light-polarization properties  in detail. The anisotropic optical emission of the CrSBr layer is evidenced by linear-polarization-resolved PL measurements. Figure~\ref{fig.LinearDependence}~(e) shows a color map of the linearly-polarized PL intensity as a function of the in-plane linear polarization angle at 3.6~K.  The  polar plots for these emissions are shown in Figure~\ref{fig.LinearDependence}~(f). All CrSBr PL peaks are strongly linearly polarized along the $\hat{b}$ axis, which evidences the anisotropic electronic structure of CrSBr, as expected. Remarkably, a clear dependence of the PL intensity on the in-plane polarization angle is observed for both, the MoSe$_2$ exciton [black squares in Fig.~\ref{fig.LinearDependence}~(f)] and trion [red circles in Fig.~\ref{fig.LinearDependence}~(f)] emissions. This result indicates that the MoSe$_2$ has acquired a linear-polarization component along the $\hat{a}$ axis probably due to magnetic proximity or photonic effects due to the linear dichroism of CrSBr. 

We have also measured the PL for different magnetic field ($\Vec{B}$) orientations. Figures~\ref{fig.BField_Parallel+BS}~(a) and (c) show color maps of the MoSe$_2$/CrSBr  magneto-PL intensity under $\Vec{B}$ parallel to the in-plane easy ($\Vec{B} \parallel \hat{b}$) and hard axis ($\Vec{B} \parallel \hat{a}$), respectively. For $\Vec{B} \parallel \hat{b}$, the PL spectrum of CrSBr red-shifts abruptly by about 15~meV above a field of 0.375~T and is constant above 0.375~T (see also Figures~S6 and S7). This result is similar to previous magneto-optical measurements for few-layer CrSBr\cite{Wilson2021} and was explained by a spin-flip transition from AFM to FM order also observed in magnetization measurements\cite{Klein2023}. Under $\Vec{B} \parallel \hat{a}$, the PL spectrum shifts smoothly, due to the canting of the spins along $\Vec{B}$, saturating at $B=1.075$~T beyond which the PL spectrum remains unchanged.  The observed PL red shifts of CrSBr with increasing magnetic field was explained by a magnetization-dependent interlayer electronic coupling in the CrSBr material \cite{Wilson2021}.

Remarkably, we also find that the PL intensities of the MoSe$_2$ trion and exciton are correlated to the field-induced phase transition in CrSBr bulk: for $\Vec{B} \parallel \hat{b}$ an abrupt change of the PL intensity of MoSe$_2$ above the critical magnetic field of 0.375~T occurs [see Figure~\ref{fig.BField_Parallel+BS}~(b) and Figures~S6 and S7], and for $\Vec{B} \parallel \hat{a}$, a continuous decrease of both MoSe$_2$ PL intensities is present up to 1.075~T, which corresponds to the saturation of the magnetization in CrSBr. Furthermore, the relative intensity of the trion/exciton peaks (Figure~S8) also shows an abrupt change for $\Vec{B} \parallel \hat{b}$, above the critical field 0.375~T, and a continuous change up to 1.075~T  for $\Vec{B} \parallel \hat{a}$, which would indicate an increase in the doping of MoSe$_2$. This could be explained by a change of charge transfer after the magnetic-field-induced phase transition. 

These results can be rationalized by our first principles calculations of the electronic band structure shown in Figures~\ref{fig.BField_Parallel+BS}~(e-g). Not only the electronic structures of CrSBr in the AFM/FM phases are different\cite{Wilson2021,Bianchi2023PRB} but also their band alignment (type-III) with respect to MoSe$_2$ changes [see bottom panels in Figures~\ref{fig.BField_Parallel+BS}~(e-g)]. These energetic differences suggest that the charge transfer between ML MoSe$_2$ and CrSBr can be drastically altered when increasing the magnetic field because of the transition from AFM to FM phases in CrSBr. 

\begin{figure*}[htb]
\centering{\includegraphics[width=2\columnwidth]{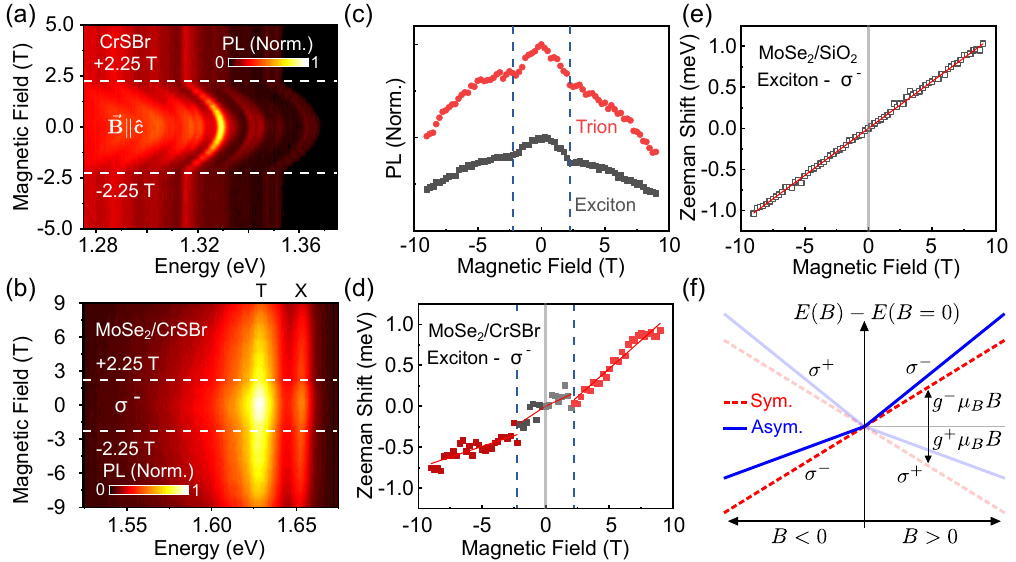}}
\caption{Color-code map of the circularly resolved PL intensity as a function of out-of-plane magnetic field for (a) CrSBr and (b) MoSe$_2$/CrSBr. The laser excitation is linearly polarized and the PL detection is $\sigma_-$ for positive magnetic field. (c) PL intensity of exciton and trion peaks of MoSe$_2$/CrSBr as a function of magnetic field. Zeeman shift for the exciton peaks in the region of (d) MoSe$_2$/CrSBr and (e) MoSe$_2$/SiO$2$. The solid lines are the fittings to the data. The extracted $g$-factors are summarized in Table~\ref{tab:gfactors}. (f) Schematic representation of symmetric (dashed lines) and asymmetric (solid lines) Zeeman shifts as function of magnetic field. The transparent lines indicate the $\sigma^+$ polarization that is not being measured.}
\label{fig.BField_Perpendicular}
\end{figure*}

Let us now turn to the magneto-PL investigations of the MoSe$_2$/CrSBr heterostructure for an out-of-plane magnetic field ($\Vec{B} \parallel \hat{c}$) under linearly-polarized excitation and $\sigma^-$ circularly-polarized PL detection as a function of $\Vec{B}$. For  CrSBr emission energies [Figure~\ref{fig.BField_Perpendicular}~(a)], a continuous red-shift of all PL peaks occurs while increasing $B$ (in absolute value) up to a saturation field of about 2.25~T, beyond which the PL peaks remains unchanged, consistent with previous reports \cite{Wilson2021}. For MoSe$_2$, the color code map of PL intensity as a function of $B$ is shown in Figure~\ref{fig.BField_Perpendicular}~(b). It also exhibits a correlation with the magnetic phase order of CrSBr. Figure~\ref{fig.BField_Perpendicular}~(c)  presents the intensities of the exciton and trion PL peaks as a function of the $B$. We observe an unusual change of the PL intensity for both, exciton and trion, in the range of -2.25 to +2.25~T, which is correlated to the magnetic-field-induced phase transition of CrSBr. In addition, we observe a blue (red) shift of PL peak positions as shown in Figure 3 (b) and (d)  with an increase of positive (negative) $B$ values, resembling the effects of the valley Zeeman splitting \cite{Macneill2015, Wang2015}.

The $B$ dependence for one particular polarization branch ($\sigma^+$ or $\sigma^-$) of the PL peak of the exciton or trion in TMDs can be written as\cite{Stier2016, Stier2018PRL, Klein2021PRR}
\begin{equation}
 E_{i}(B) = E_{i}(B=0) + g_{i}^{j}\mu_B B,
\label{eq.magneticshifts}
\end{equation}
in which $\mu_B$ is the Bohr magneton, the subindex $i = \text{X (T)}$ identifies the exciton (trion), and the superindex $j=\pm$ denotes the circular polarization $\sigma^\pm$. Equation~(\ref{eq.magneticshifts}) describes the Zeeman shift of one polarization branch (the increase or decrease depends on the sign of $g_{i}^{j}$, which is system dependent), whereas the Zeeman splitting requires the knowledge of the Zeeman shifts for each polarization. The total $g$-factor  that modulates the Zeeman splitting is then given by $g_i = g_i^{+} - g_i^{-}$. In pristine monolayer TMDs, $g_\text{X}^{+} \sim -2$ and $g_\text{X}^{-} \sim 2$ (directly related to the angular momenta of the valence and conduction band states at the K valleys involved in the exciton transition\cite{Wozniak2020, FariaJunior2022NJP}), leading to a total  $g$-factor of $g_\text{X} \sim -4$. Furthermore, time reversal symmetry connects the $g$-factors obtained at positive and negative magnetic fields via $g_i^{+}(B>0) = -g_i^{-}(B<0)$, allowing us to recover the Zeeman shift of the $\sigma^+$ branch by measuring the $\sigma^-$ branch at negative magnetic fields.

\begin{table*}[htb]
\caption{Exciton and trion $g$-factors for $B>0$. The $g$-factors for $\sigma^+$ were obtained via $g_i^{+} = -g_i^{-}(B<0)$ and the total $g$-factor is given by $g_i = g_i^{+} - g_i^{-}$. In the AFM phase, acessed by small fields, the Zeeman shifts approach the spectral resolution of the system, resulting in higher error bars for the obtained values. Nevertheless, the errors are still smaller than the extracted $g$-factors and allow us to unambiguously identify the asymmetric signatures.}
\begin{tabular}{cccccccc}
\hline 
\hline 
 &  &  &  &  &  &  & \tabularnewline [-3mm]
 &  &  & MoSe$_{2}$/SiO$_{2}$ &  & \multicolumn{3}{c}{MoSe$_{2}$/CrSBr } \tabularnewline [2mm]
 &  &  &  &  & AFM &  & FM\tabularnewline [2mm]
\hline 
&  &  &  &  &  &  & \tabularnewline [-3mm]
        & $g_{\text{X}}^{+}$ &  & $-1.98\pm0.05$ &  & $-1.75\pm0.39$ &  & $-0.90\pm0.05$\tabularnewline [2mm]
Exciton & $g_{\text{X}}^{-}$ &  & $\;\;1.97\pm0.05$ &  & $\;\;1.25\pm0.44$ &  & $\;\;2.37\pm0.05$\tabularnewline [2mm]
        & $g_{\text{X}}$ &  & $-4.0\pm0.1$ &  & $-3.0\pm0.8$ &  & $-3.3\pm0.1$\tabularnewline [2mm]
\hline 
&  &  &  &  &  &  & \tabularnewline [-3mm]
      & $g_{\text{T}}^{+}$ &  & $-2.10\pm0.05$ &  & $-1.33\pm0.18$ &  & $-1.42\pm0.06$\tabularnewline [2mm]
Trion & $g_{\text{T}}^{-}$ &  & $\;\;2.14\pm0.05$ &  & $\;\;2.25\pm0.46$ &  & $\;\;1.84\pm0.06$\tabularnewline [2mm]
      & $g_{\text{T}}$ &  & $-4.2\pm0.1$ &  & $-3.6\pm0.6$ &  & $-3.3\pm0.1$\tabularnewline [1mm]
\hline 
\hline
\end{tabular}
\label{tab:gfactors}
\end{table*}

The excitonic Zeeman shift obtained for the MoSe$_2$/CrSBr heterostructure is displayed in Figure~\ref{fig.BField_Perpendicular}~(d). As a reference, we have also measured the magneto-PL in the MoSe$_2$/SiO$_2$ region of the sample [see Figure~\ref{fig.BField_Perpendicular}~(e)]. The Zeeman shifts of the trion peaks are presented in  Figure~S10  and follow closely the excitonic features. Our results reveal an intriguing asymmetric signature in the Zeeman shift of the MoSe$_2$ exciton within the MoSe$_2$/CrSBr heterostructure, while the MoSe$_2$/SiO$_2$ system displays a symmetric response. These findings point to an asymmetric coupling between the MoSe$_2$ valleys and the CrSBr bands, which is dependent on the magnetic ordering. Exploiting time reversal symmetry allows us to extract distinct $g_i^{+}$ and $g_i^{-}$ values for each magnetic phase at positive and negative magnetic fields. In Figure~\ref{fig.BField_Perpendicular}~(f), we present a schematic representation of the symmetric and asymmetric Zeeman shifts, summarizing the observed features of Figures~\ref{fig.BField_Perpendicular}~(d,e). The obtained $g$-factors for excitons and trions are summarized in Table \ref{tab:gfactors}. Particularly, for the excitons in MoSe$_2$/SiO$_2$, we extract $\left|g_{\text{X,T}}^{+}\right|=\left|g_{\text{X,T}}^{-}\right|$ leading to a total $g$-factor of $\sim-4.0$, consistent with theoretical\cite{Wozniak2020, Deilmann2020PRL, Xuan2020PRR, FariaJunior2022NJP} and experimental values, reported in the literature for MoSe$_2$/SiO$_2$ or hBN/MoSe$_2$/hBN\cite{Li2014, Macneill2015, Wang2015, Mitioglu2016, Arora2018, Koperski2018, Goryca2019, Robert2020, Gobato2022, Covre2022}. 
For the MoSe$_2$ excitons on CrSBr, $\left|g_{\text{X,T}}^{+}\right|$ is distinctly different from $\left|g_{\text{X,T}}^{-}\right|$ and the total $g$-factor is less negative than the typical values of $-4$ in pristine MoSe$_2$.
Our study uncovers notable variations in the $g$-factors of the MoSe$_2$ exciton and trion when the CrSBr undergoes transitions between the AFM and FM phases, revealing an asymmetric coupling between the spin-valley properties of MoSe$_2$ and the magnetic ordering of CrSBr. These distinct $g$-factors provide valuable insights into the intricate interplay between electronic and magnetic degrees of freedom, underscoring the importance of considering the magnetic state of CrSBr in understanding the behavior of excitonic systems in this heterostructure. The changes in the magnitude of the $g$-factors are consistent with proximity effects due to the hybridization between the layers, as previously demonstrated in MoSe$_2$/WSe$_2$\cite{Wozniak2020,FariaJunior2023Nanomat}, WSe$_2$/CrI$_3$\cite{Heissenbuttel2021NL}, and WS$_2$/graphene systems\cite{FariaJunior2023TDM}. A systematic analysis of the microscopic features behind the asymmetric $g$-factors is beyond the scope of the current manuscript; however, we point out that asymmetric signatures in valley Zeeman splitting have recently been observed in MoSe$_2$/CrBr$_3$\cite{Choi2022_MoSe2-CrBr3} heterostructures at zero magnetic field. In these systems the magnetic moments in CrBr$_3$ point in the out-of-plane direction and act already as an external magnetic field. Here, the magnetic moments of CrSBr are oriented in-plane and therefore the asymmetric coupling is manifested once we apply an external magnetic field. The asymmetric Zeeman shifts do not necessarily require a magnetic material but can also be present in systems where valence bands are mixed\cite{Tedeschi2019PRB}.

Furthermore, we have also measured the linear polarization of the PL of the heterostructure [see Figure~S4~(f)]. We find that the angle dependence and relative intensity of the trion/exciton of the MoSe$_2$ PL are clearly modified as compared to 0~T. The observed anisotropy of the relative intensities of MoSe$_2$ trion/exciton could be explained by an anisotropic band structure of the heterostructure due to proximity effects.

\begin{figure*}[ht]
\centering{\includegraphics[width=2.0\columnwidth]{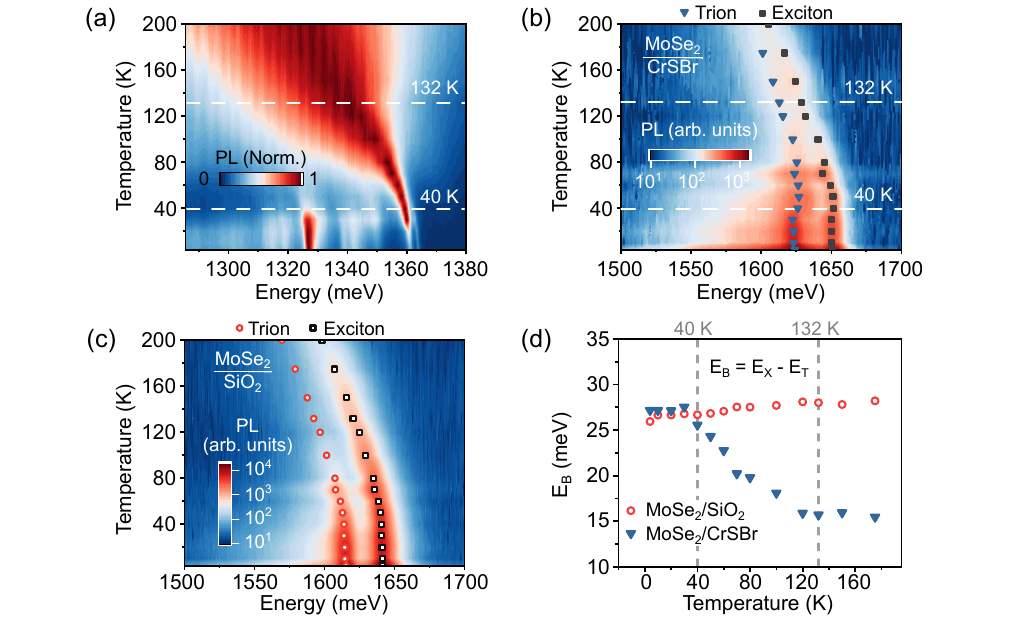}}
\caption{(a) Color-code map of PL intensity
as a function of temperature for the PL peaks
of the CrSBr. The highest dot-dashed line indicates the CrSBr Néel temperature transition at 132~K, and the lower one at 40~K indicates the temperature where sharpening of the peaks appears. (b,c) Color-code maps of the PL intensity as a function of the temperature for the exciton and trion peaks from (b) MoSe$_2$/CrSBr and (c) MoSe$_2$/SiO$_2$. (d) Trion binding energy extracted from the data shown in (c) and (d).}\label{fig.TemperatureDependence}
\end{figure*}

 We now analyze the temperature dependence of the PL data shown in Figure~\ref{fig.TemperatureDependence}. For CrSBr, a blue shift of the PL band is observed  with decreasing temperature, which is accompanied by a change in the peak shape around the magnetic phase transition ($T_N$ around 132~K). In addition, at 40~K, sharp peaks appear below 1360~meV, together with a clear enhancement of the PL intensity of the peak at around 1330~meV. A clear correlation between the emission peaks and phase transitions in CrSBr is thus present.

Important changes are also observed for the PL of MoSe$_2$. At higher temperatures  ($T_N$ > 132~K),  the trion binding energy of MoSe$_2$/CrSBr is much lower than that of MoSe$_2$/SiO$_2$, probably due to different dielectric constant values of CrSBr and SiO$_2$. Remarkably, we find an anomalous temperature dependence of the exciton and trion peak positions for MoSe$_2$/CrSBr. This is visualized in Figure~\ref{fig.TemperatureDependence} (d) (see also Figures S11 and S12), where we plot the extracted trion binding energies versus temperature. The MoSe$_2$/CrSBr trion binding energy increases with decreasing temperature between the magnetic phase transitions, while it stabilizes above $T_N$ and below 40~K.  A similar anomaly was observed in the temperature dependence of excitons in the MoSe$_2$/MnPSe$_3$ heterostructure near $T_N$, and was associated to a coupling of MoSe$_2$ excitons to magnons in MnPSe$_3$\cite{Onga2020}. In our heterostructure, MoSe$_2$ excitons may also couple to the (incoherent) magnons \cite{Dirnberger2023} of CrSBr at non-zero temperatures. The impact of these magnons on the CrSBr band structure has not yet been studied in detail, but it is expected that magnon-induced changes will affect both the charge transfer between MoSe$_2$ and CrSBr as well as the dielectric screening experienced by the excitons in MoSe$_2$. Both phenomena may contribute to the exciton/trion temperature dependence \cite{VanTuan2018,Florian2018,Zhumagulov2020,Lin2014}.
However, further studies will be necessary to understand in more detail this experimental result.

In summary, we have measured the linearly and circularly polarized PL on  MoSe$_2$/CrSBr heterostructures under magnetic fields up to 9~T oriented along the different crystallographic axes of CrSBr . The results show that the valley and excitonic properties (intensity, energy position, and $g$-factors) of monolayer MoSe$_2$ are strongly influenced by the magnetic order of a CrSBr substrate. For all magnetic field orientations we found that the MoSe$_2$ PL intensity is sensitive to the magnetic ordering of the CrSBr. We predict a type-III band alignment for MoSe$_2$/CrSBr which can account for the observed correlation of  MoSe$_2$ PL intensity with the magnetic induced phase transition of CrSBr. For out-of-plane magnetic fields, a clear asymmetric Zeeman shift is observed for MoSe$_2$/CrSBr.  Furthermore, we observe an anomalous behaviour of the trion binding energy as a function of temperature. The binding energy is considerably low at high temperatures and increases below $T_N$. In general, our results are explained by asymmetric magnetic proximity, charge transfer, exciton/trion magnon coupling and dielectric anomalies of the 2D antiferromagnetic material. Our findings offer a unique insight into  the interplay of proximity effects and charge transfer in antiferromagnetic-nonmagnetic interfaces that modify the exciton and valley properties of 2D TMDs.

\begin{acknowledgement}

This work was supported by  Fundação de Amparo a Pesquisa do Estado de São Paulo (FAPESP) (grants 22/08329-0 and 23/01313-4) and by the Brazilian Council for Research (CNPq) (grant 311678/2020-3). CSB acknowledges the financial support of CAPES fellowship. TSG and HvdZ received funding from European Union Horizon 2020 research and innovation program under grant agreement No. 863098 (SPRING). PEFJ, KZ, CS, and JF acknowledge the financial support of the Deutsche Forschungsgemeinschaft (DFG, German Research Foundation) SFB 1277 (Project-ID 314695032, projects B05, B07 and B11), SPP 2244 (Project No. 443416183, SCHU1171/10), and of the European Union Horizon 2020 Research and Innovation Program under Contract No. 881603 (Graphene Flagship).
YGG and HvdZ acknowledge support from the Fapesp-SPRINT project  (grant 22/00419-0). K.W. and T.T. acknowledge support from the JSPS KAKENHI (Grant Numbers 21H05233 and 23H02052) and World Premier International Research Center Initiative (WPI), MEXT, Japan. S.M.-V. acknowledges the European Commission for a Marie Sklodowska–Curie individual fellowship No. 101103355 - SPIN-2D-LIGHT.  J.I.A. acknowledges support from the European Union’s Horizon 2020 research and innovation programme for a Marie Sklodowska–Curie individual fellowship No. 101027187-PCSV.
F.D. acknowledges financial support from Alexey Chernikov and the W\"urzburg-Dresden Cluster of Excellence on Complexity and Topology in Quantum Matter ct.qmat  (EXC 2147, Project-ID 390858490).

\end{acknowledgement}

\begin{suppinfo}

The following files are available free of charge.
\begin{itemize}
  \item SI: Sample preparation, experimental methods and complementary PL results.  Details on the first principles calculations. (PDF)
 
\end{itemize}

\end{suppinfo}

\bibliography{References}

\end{document}